\documentclass[structabstract]{aa}  
\usepackage{savesym}
\usepackage{amsmath}
\savesymbol{iint}
\usepackage{graphicx}
\usepackage{xspace}
\usepackage{booktabs}
\usepackage[varg]{txfonts}
\usepackage{natbib,twoopt}
\usepackage[breaklinks=f]{hyperref}
\usepackage{epsfig}
\usepackage{txfonts}
\usepackage{amssymb}
\usepackage{color}
\restoresymbol{TXF}{iint}
\bibpunct{(}{)}{;}{a}{}{,} 
 
\newcommandtwoopt{\citeads}[3][][]{\href{http://adsabs.harvard.edu/abs/#3}%
{\citealp[#1][#2]{#3}}}
\newcommandtwoopt{\citepads}[3][][]{\href{http://adsabs.harvard.edu/abs/#3}%
{\citep[#1][#2]{#3}}}
\newcommandtwoopt{\citetads}[3][][]{\href{http://adsabs.harvard.edu/abs/#3}%
{\citet[#1][#2]{#3}}} 
\newcommandtwoopt{\citeyearads}[3][][]%
{\href{http://adsabs.harvard.edu/abs/#3}{\citeyear[#1][#2]{#3}}}

 \def\teff{$T_\mathrm{eff}$}                 
 \def\vmic{$v_\mathrm{mic}$}                 
 \def\vmac{$v_\mathrm{mac}$}                 
 \def\vsini{\hbox{$v$\,sin\,$i_{\star}$}}    
 \def\ms{\hbox{\,m\,s$^{-1}$}}               
 \def\m2s2{\hbox{\,m$^{2}$\,s$^{-2}$}}       
 \def\kms{\hbox{\,km\,s$^{-1}$}}             
 \def\cms2{\hbox{\,cm\,s$^{-2}$}}           
 \def\gcm3{\hbox{\,g\,cm$^{-3}$}}            
 \def\Msun{\hbox{$\mathrm{M}_{\odot}$}}      
 \def\Rsun{\hbox{$\mathrm{R}_{\odot}$}}      
\begin{document}

\title{Mass determination of K2-19b and K2-19c \\ from radial velocities and transit timing variations}

   \author{D.~Nespral\inst{\ref{IAC},\ref{lalaguna}},
           D.~Gandolfi\inst{\ref{Torino},\ref{LSW}},
           H.\,J. Deeg\inst{\ref{IAC},\ref{lalaguna}},
           L.~Borsato\inst{\ref{Padova}},
           M.\,C.\,V Fridlund\inst{\ref{Onsala},\ref{Leiden}},
           O.~Barrag\'an\inst{\ref{Torino}}, 
           R.~Alonso\inst{\ref{IAC},\ref{lalaguna}}
           S.~Grziwa\inst{\ref{Koln}},
           J.~Korth\inst{\ref{Koln}},
           S.~Albrecht\inst{\ref{SAC}},
           J.~Cabrera\inst{\ref{DLR}},
           Sz.~Csizmadia\inst{\ref{DLR}},
           G.~Nowak\inst{\ref{IAC},\ref{lalaguna}},
           T.~Kuutma\inst{\ref{NOT}},
           J.~Saario\inst{\ref{NOT}},
           P.~Eigm\"uller\inst{\ref{DLR}},
           A.~Erikson\inst{\ref{DLR}},
           E.\,W.~Guenther\inst{\ref{TLS}},
           A.\,P.~Hatzes\inst{\ref{TLS}},
           P.~Monta\~n\'es Rodr\'iguez\inst{\ref{IAC},\ref{lalaguna}},
           E.~Palle\inst{\ref{IAC},\ref{lalaguna}},
           M.~P\"atzold\inst{\ref{Koln}},
           J.~Prieto-Arranz\inst{\ref{IAC},\ref{lalaguna}},
           H.~Rauer\inst{\ref{DLR},\ref{TUBerlin}},
           D.~Sebastian\inst{\ref{TLS}}
           }

\institute{Instituto de Astrofísica de Canarias, C/ Vía Láctea s/n, E-38205, La Laguna, Tenerife, Spain\label{IAC}
\and Dpto. Astrofísica Universidad de La Laguna, Tenerife, Spain \label{lalaguna} 
\and Dipartimento di Fisica, Universit\'a degli Studi di Torino, via Pietro Giuria 1, I-10125, Torino, Italy\label{Torino}      
\and Landessternwarte K\"onigstuhl, Zentrum f\"ur Astronomie der Universit\"at Heidelberg, K\"onigstuhl 12, D-69117 Heidelberg, Germany\label{LSW}
\and Dipartimento di Fisca e Astronomia, Universit\'a degli Studi di Padova, Via Marzolo 8, I-35131, Padova, Italy\label{Padova}
\and Leiden Observatory, University of Leiden, PO Box 9513, 2300 RA, Leiden, The Netherlands\label{Leiden}
\and Department of Earth and Space Sciences, Chalmers University of Technology, Onsala Space Observatory, 439 92 Onsala, Sweden\label{Onsala}
\and Rheinisches Institut f\"ur Umweltforschung, Abteilung Planetenforschung an der Universit\"at zu K\"oln, Aachener Strasse 209, 50931 K\"oln, Germany\label{Koln}
\and Stellar Astrophysics Centre, Deparment of Physics and Astronomy, Aarhus University, Ny Munkegrade 120, DK-8000 Aarhus C, Denmark\label{SAC}
\and Institute of Planetary Research, German Aerospace Center, Rutherfordstrasse 2, 12489 Berlin, Germany\label{DLR}
\and Th\"uringer Landessternwarte Tautenburg, Sternwarte 5, D-07778 Tautenberg, Germany\label{TLS}
\and Nordic Optical Telescope, Apartado 474, 38700 Santa Cruz de La Palma, Spain\label{NOT}
\and Center for Astronomy and Astrophysics, TU Berlin, Hardenbergstr. 36, 10623 Berlin, Germany\label{TUBerlin}
}
        
\date{Received  / Accepted }

\abstract{We present FIES@NOT, HARPS-N@TNG, and HARPS@ESO-3.6m radial velocity follow-up observations of \object{K2-19}, a compact planetary system hosting three planets, of which the two larger ones, namely \object{K2-19b} and \object{K2-19c}, are close to the 3:2 mean motion resonance. An analysis considering only the radial velocity measurements detects K2-19b, the largest and most massive planet in the system, with a mass of $54.8\pm7.5$~M${_\oplus}$ and provides a marginal detection of K2-19c, with a mass of M$_\mathrm{c}$\,=\,$5.9^{+7.6}_{-4.3}$\,M$_\oplus$. We also used the \texttt{TRADES} code to simultaneously model both our RV measurements and the existing transit-timing measurements. We derived a mass of $54.4\pm8.9$~M${_\oplus}$ for K2-19b and of $7.5^{+3.0}_{-1.4}$~M${_\oplus}$ for K2-19c. A prior K2-19b mass estimated by \citet{Barros2015}, based principally on a photodynamical analysis of K2-19's light-curve, is consistent with both analysis, our combined TTV and RV analysis, and with our analysis based purely on RV measurements. 
Differences remain mainly in the errors of the more lightweight planet, driven likely by the limited precision of the RV measurements and possibly some yet unrecognized systematics.}

               
\titlerunning{Mass determination of K2-19b and K2-19c from RVs and TTVs}
\authorrunning{Nespral, Gandolfi, Deeg, et al.}

   \maketitle

\section{Introduction}
 
Planets in mean-motion resonances (MMR) or commensurabilities have orbital period ratios that are close to integer values. Several MMR are found in the solar system and are regarded as ``tale tellers'' of its dynamical evolution. For instance, the Neptune/Pluto 3:2 MMR is believed to be the result of an outward migration of Neptune \citep{Petrovich2013}. According to the Grand Tack model, Jupiter and Saturn got trapped in a 3:2 (or 2:1) resonance in the early phases of the solar system formation. This would have halted and inverted the inward migration of Jupiter at $\sim$1.5\,AU, shaping the architecture of the inner terrestrial planets as we know it today \citep[see, e.g.,][]{Pierens2014}.

Exoplanets can also be driven into resonant configurations through dissipative mechanisms that can change the energy of their orbits and, thus, the corresponding semi-major axis \citep{Plavchan2013}. The MMRs most frequently found in exoplanetary systems are the 2:1 and 3:2 resonances, though others might also exist \citep[see, e.g.,][]{Holman2010,Fabrycky2012,Petrovich2013,Fabrycky2014}. Current scenarios of planet formation allow for the formation of planets at any orbital radii. Therefore, it is believed that resonant configurations did not come into place during the formation of planets, but are rather the outcome of the dynamical evolution of planetary systems. MMRs can thus provide precious insights into the evolution history of planetary systems \citep{Kley2010}.

\object{K2-19} (also known as \object{EPIC\,201505350}) is a V=13\,mag late-type star observed by the K2 space mission during its Campaign~1. It hosts three transiting planets, with the orbital periods of the two larger planets being close to the 3:2 mean motion resonance: \object{K2-19b}, a sub-Saturn-size planet with an orbital period of $\sim$7.9 days, and \object{K2-19c}, a Neptune-size planet with an orbital period of $\sim$11.9 days. The two planets perturb each other causing transit timing variations (TTVs) that are visible both in the K2 data \citep{Armstrong2015} and in ground-based transit observations \citep{Narita2015,Barros2015}. A small inner planet with a radius of $1.14\pm0.13$\,R${_\oplus}$, namely \object{K2-19d}, was recently found to transit the star every $\sim$2.5 days \citep{Sinukoff2016}. We do not consider K2-19d in the following analysis, given its expected small mass and the absence of any reported transit timing variations.

Systems such as K2-19 are precious and unique laboratories to study planet formation, migration, and evolution \citep{Armstrong2015}, as their orbital architectures imply a common inward migration scenario for the resonant planets \citep{Naoz2015}. There seems to be a lack of short period gas giants in 2:1 and 3:2 MMRs, which is likely due to dynamical instability of these systems \citep{Narita2015}. K2-19b is to date the only gas giant planet with an orbital period shorter than 50 days known to be in a 3:2 MMR. In addition, K2-19 is a unique system compared to other resonant systems, because the inner planet K2-19b is larger (and likely more massive) than the outer K2-19c, whereas outer planets in 2:1 or 3:2 MMRs tend to be larger and more massive than the inner ones. An accurate mass determination of K2-19b would be an important piece of the puzzle for understanding the dynamic of such systems \citep{Ogihara2013}.

Several attempts have been made to determine the masses of K2-19b and K2-19c. \citet{Armstrong2015} combined K2 data with ground-based transit photometry of K2-19b and used the observed TTVs to put some constraints on the mass of the two planets.  
\citet{Barros2015} used a more sophisticated approach to derive the masses of the planets, based on a photo-dynamical model that considers transit timings and durations from transits observed by the K2 mission as well as two additional K2-19b transits observed from ground. They included also radial velocities obtained with SOPHIE at the OHP-1.9-m telescope in their analysis, although they realized that the precision of these RVs prevented the detection of the Doppler reflex motion induced by the planets. The photo-dynamical approach employed by \citet{Barros2015} models the data with an n-body dynamical integrator that takes into account the gravitational interactions of all components and derives the corresponding transit timings. Furthermore, the photo-dynamical model was executed as part of a Monte-Carlo Markov chain (MCMC) that in principle permits reliable estimates of the planet parameters, given the uncertainties of the TTVs and other input parameters. They found a mass of M$_\mathrm{b}$=44$\pm$12~M${_\oplus}$ for K2-19b, and M$_\mathrm{c}$=15.9$\pm$7~M${_\oplus}$ for K2-19c. Shortly afterwards, \citet{Narita2015} presented additional ground-based transit photometry of K2-19b and modeled the observed TTVs using the ``synodic chopping'' formulae given by \citep{Deck2015}. They found two possible solutions that are positioned above and below the 3:2 MMR. Despite the degeneracy of their solution, they estimated the mass of the outer planet K2-19c to be M$_\mathrm{c}$$\sim$20~M${_\oplus}$, the latter in agreement with \citet{Barros2015}. Although \citet{Narita2015} did not include the transit timings from \citet{Barros2015}, the follow-up observations of both groups were taken at similar dates and the derived TTVs agree within the error bars. 
While the photo-dynamical approach by \citet{Barros2015} is in principle reliable and independent of any simplifying assumptions, we note that their analysis is based on an MCMC of no more than 3\,500 independents points -- apparently limited by the computing requirements of the complex model calculations -- and that MCMC in highly non-linear situations -- definitively the case for models where TTVs are input parameters -- may easily get stuck around non-optimum solutions. \citet{Barros2015} in their Sect.~5.2 compare their results to a simplified analysis based solely on the K2 light-curve, giving results that are in agreement but with wider posterior parameter distributions. We can therefore not derive any conclusion if the presence or absence of their RV measurements had any effect in their mass-determinations. We also note the work of \citet{Weiss2014} who studied the TTV-derived masses of 65 small exoplanets (R$_\mathrm{p}$\,$\le$\,4\,R${_\oplus}$) and compared them with those derived with RV measurements. They found that masses from TTVs are systematically lower than masses from RVs. An independent verification of the masses of K2-19b and K2-19c from \citet{Barros2015} and \citet{Narita2015} is therefore desirable.

Here we present a high-resolution spectroscopic follow-up of K2-19 and new estimates of the masses of K2-19b and K2-19c. The paper is organized as follows: in Sect.~\ref{RVs} we describe the observations; in Sect.~\ref{StellarParameters} we present the spectral analysis and the properties of the host star; in Sect.~\ref{RV_Analysis} and \ref{RV_TTV_Analysis} we report on our data analysis. We present and discuss our results in Sect.~\ref{Discussion}.

\section{High-resolution spectroscopic follow-up}
\label{RVs}

We used the FIbre-fed \'Echelle Spectrograph \citep[FIES;][]{Frandsen1999,Telting2014} mounted at the 2.56-m Nordic Optical Telescope (NOT) at Roque de los Muchachos Observatory (La Palma, Spain) to collect 10 high-precision RVs of K2-19. The observations were carried out between January 2015 and January 2016 as part of the CAT and OPTICON observing programs 109-MULTIPLE-2/14B, 35-MULTIPLE-2/15B, and 15B/064. We used the FIES \emph{high-res} mode, which provides a resolving power of R\,$\approx$\,67\,000 in the spectral range 3600\,--\,7400\,\AA. Following the observing strategy described in \citet{Buchhave2010} and \citet{Gandolfi2015}, we took 3 consecutive exposures of 900-1200 seconds per observation epoch -- to remove cosmic ray hits -- and acquired long-exposed (T$_\mathrm{exp}$\,$\approx$\,35 seconds) ThAr spectra immediately before and after the three sub-exposures -- to trace the RV drift of the instrument. We reduced the data using standard IRAF and IDL routines, which include bias subtraction, flat fielding, order tracing and extraction, and wavelength calibration. The signal-to-noise ratio (SNR) of the extracted spectra is about $\sim$20-25 per pixel at 5500~\AA. Radial velocity measurements were derived via SNR-weighted, multi-order, cross-correlation with the RV standard star \object{HD\,50692} -- observed with the same instrument set-up as K2-19.

\begin{table}
\begin{center}
\caption{FIES, HARPS-N, and HARPS measurements of K2-19.}
\begin{tabular}{lccc}
\hline
\hline
\noalign{\smallskip}
BJD$_\mathrm{TDB}$ &   RV   & $\sigma_\mathrm{RV}$  & CCF Bis. Span \\
 -2\,450\,000          & [\kms] &        [\kms]         &    [\kms]  \\
\noalign{\smallskip}
\hline
\noalign{\smallskip}
\multicolumn{4}{l}{FIES} \\
\noalign{\smallskip}
7045.70173	& 7.1893  & 0.0148  & $-$0.0311 \\
7049.75170	& 7.2181  & 0.0156  & $-$0.0372 \\
7051.70510	& 7.1728  & 0.0147  & $-$0.0273 \\
7053.73321	& 7.1927  & 0.0149  & ~\,\,0.0096 \\
7054.73211	& 7.1820  & 0.0096  & $-$0.0059 \\
7065.66244  & 7.2207  & 0.0105  & $-$0.0142 \\
7392.75475	& 7.1791  & 0.0167  & ~\,\,0.0084 \\
7394.74622	& 7.1772  & 0.0163  & $-$0.0136  \\
7395.67480  & 7.1938  & 0.0140  & $-$0.0213  \\
7398.72585	& 7.2113  & 0.0130  & $-$0.0436  \\
\noalign{\smallskip}
\hline
\noalign{\smallskip}
\multicolumn{4}{l}{HARPS-N} \\
\noalign{\smallskip}
7064.62294  & 7.3433  & 0.0051  & $-$0.0259  \\
7064.64366  & 7.3378  & 0.0060  & $-$0.0231  \\
7142.43784  & 7.3199  & 0.0054  & $-$0.0318  \\
7370.77006  & 7.2989  & 0.0050  &  ~\,\,0.0014 \\
7370.79235  & 7.2974  & 0.0051  & $-$0.0216  \\
7372.77083  & 7.3090  & 0.0087  & $-$0.0116  \\
7372.78621  & 7.3189  & 0.0065  & $-$0.0256  \\
7448.55938  & 7.2932  & 0.0090  & $-$0.0028  \\
7492.48547  & 7.3296  & 0.0034  & $-$0.0140  \\
\noalign{\smallskip}
\hline
\noalign{\smallskip}
\multicolumn{4}{l}{HARPS} \\
\noalign{\smallskip}
7509.56689 & 7.3326 & 0.0089 &  ~\,\,0.0141 \\
7511.57666 & 7.3117 & 0.0028 &    $-$0.0158 \\
7512.56990 & 7.3051 & 0.0122 &    $-$0.0099 \\
\noalign{\smallskip}
\hline
\end{tabular}
\label{radialvelocity}
\end{center}
\end{table}

We also acquired 9 high-resolution spectra (R$\approx$115\,000) with the HARPS-N spectrograph \citep{Cosentino2012} based on the 3.58-m Telescopio Nazionale Galileo (TNG) at Roque de los Muchachos Observatory (La Palma, Spain). The observations were performed between February 2015 and April 2016 as part of the same observing programs as on FIES. For each observation epoch, we acquired two consecutive exposures of 1800 seconds, excepted for the last epoch when a single exposure of 3600 seconds has been taken. The extracted spectra have a SNR per pixel of $\sim$15-23 at 5500~\AA. We monitored the Moon background light using the second fibre and reduced the HARPS-N data with the HARPS-N pipeline. Radial velocities were extracted by cross-correlation with a G2 numerical mask \citep{Baranne96,Pepe02}.

Finally, we collected 5 high-resolution spectra  (R$\approx$115\,000) with the HARPS spectrograph \citep{Mayor2003} attached at the ESO-3.6-m telescope at La Silla Observatory (Chile). The observations were performed between April and May 2016, as part of the ESO program 097.C-0948. The exposure time was set to 3000--3600 sec, leading to a SNR of $\sim$13-35 per pixel at 5500~\AA\ on the the extracted spectra. We used the second fiber to monitor the sky background and reduced the data with the on-line HARSP pipeline. Radial velocities were extracted by cross-correlation with a G2 numerical mask.

The FIES, HARPS-N, and HARPS-S RV measurements and their 1-$\sigma$ error bars are listed in Table~\ref{radialvelocity}, along with the barycentric Julian date in barycentric dynamical time \citep[BJD$_\mathrm{TDB}$, see, e.g.,][]{Eastman2010} and the cross-correlation function (CCF) bisector spans. We rejected two out of five HARPS RVs, owing to a technical problem occurred during the observations. These measurements are not listed in Table~\ref{radialvelocity}. The FIES and HARPS-N RVs show a possible anti-correlation with the respective CCF bisector spans, the linear Pearson correlation coefficients being $-$0.42 and $-$0.62, respectively. We followed the method described in \citet{Loyd2014} to account for the uncertainties of our measurements and quantitatively assess the significance of the possible anti-correlation. We found that the probability that the FIES and HARPS-N measurements are uncorrelated in light of their uncertainties is higher than about 48\,\% and 24\,\%, respectively. We therefore concluded that there is no significant correlation between the FIES and HARPS-N RVs and the respective CCF bisector spans.

\section{Stellar parameters}
\label{StellarParameters}

We derived the fundamental spectroscopic parameters of K2-19 from the co-added FIES and HARPS-N spectra. Both data have a signal-to-noise ratio (SNR) of $\sim$55 per pixel at 5500\,\AA. We used the Spectroscopy Made Easy (\texttt{SME}) package \citep{ValentiPiskunov96} along with \texttt{ATLAS\,9} model atmospheres \citep{Castelli04}. We fixed the microturbulent \vmic\ and macroturbulent \vmac\ velocities to the values given by the calibration equations of \citet{Bruntt10} and \cite{Doyle14}, respectively. The effective temperature \teff\ was estimated by fitting synthetic line profiles to the observed wings of the H$\alpha$ and H$\beta$ lines. The surface gravity $\log\,g_{\star}$ was mainly derived by analyzing strong Ca\,I lines between 6100 and 6440~\AA. We measured the projected rotational velocity \vsini\ fitting the spectral profiles of several unblended metal lines. The FIES and HARPS-N co-added spectra provide consistent results well within the 1-$\sigma$ error bars. Our final adopted values for \teff, $\log g_{\star}$, [Fe/H], and \vsini\ are the weighted means of the values estimated by the FIES and HARPS-N spectra. Based on the \citet{Straizys81}'s calibration for dwarf stars, our estimates of the spectroscopic parameters translate into a K0\,V spectral type. We estimated stellar mass and radius using our values for $T_{\mathrm{eff}}$, $\log g_{\star}$, $[\mathrm{Fe}/\mathrm{H}]$ and the relationship between these parameters and M$_{\star}$, R$_{\star}$, as given by \cite{Torres10}. Results are given in Table~\ref{ParameterTable}; our values for the mass of M$_\star$\,=\,0.918$\pm$0.064~M$_\sun$ and the radius of R$_\star$\,=\,0.881$\pm$0.111~R$_\sun$ are about 3\% smaller than those derived by \citet{Barros2015}, but in agreement within error-bars.

The K2 light-curve of K2-19 shows periodic and quasi-periodic photometric variability with a peak-to-peak amplitude of about 1.2\,\% \citep{Armstrong2015}. Given the spectral type of the star, the observed variability is very likely ascribable to magnetic active regions (mainly Sun-like spots) carried around by stellar rotation. We measured the rotation period of the star using the {\bf auto correlation function } \citep[ACF; see, e.g.,][]{McQuillan13} applied to the K2 light-curve of K2-19. We measured a rotation period of $P_\mathrm{rot}$=$20.54\pm0.30$ days (Table~\ref{ParameterTable}). The fast Fourier transform of the light-curve shows also a significant peak at about 20.5 days (SNR=70), in agreement with the value found by the ACF. A consistent value has also been found by \citet{Armstrong2015}. Our estimates of the rotation period and stellar radius imply a maximum value of the projected rotation velocity of $2.17\pm0.27$ \kms, which agrees within $\sim$2-$\sigma$ with the spectroscopically derived value of $3.0\pm0.5$ \kms.

\section{RV data analysis}
\label{RV_Analysis}
 
 \begin{figure}
\centering
\includegraphics[width=\linewidth]{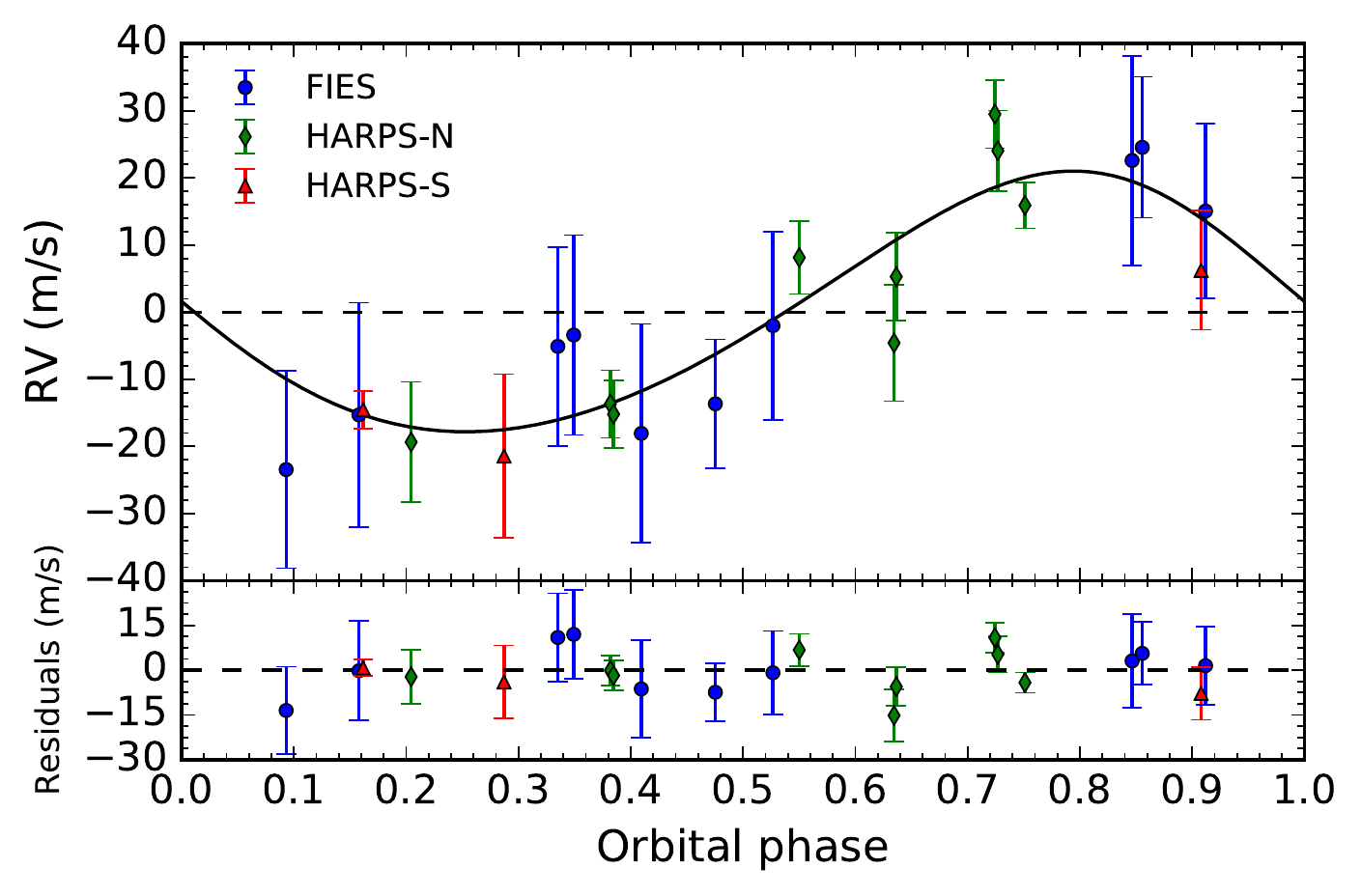}
\includegraphics[width=\linewidth]{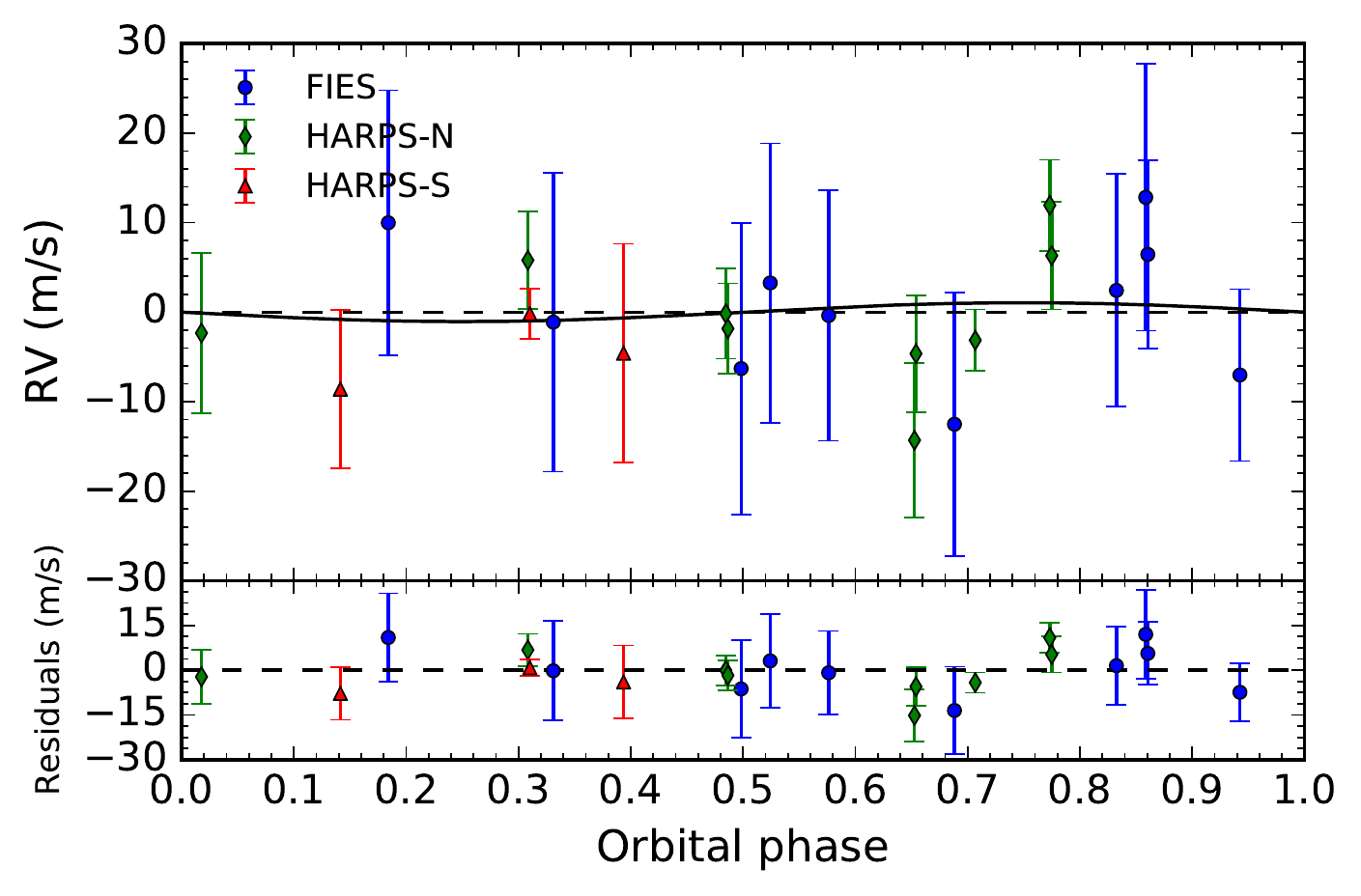}
\label{RV_Curve}
\caption{ FIES (blue circles), HARPS-N (green diamonds) HARPS-S (red triangles) RV measurements of K2-19 and Keplerian fits (solid line), phase folded to the orbital period and time of first transit of K2-19b (upper figure) and K2-19c (lower figure). For K2-19c, the fitted RVs from K2-19b have been removed. All RVs, fits and residuals (in smaller sub-panels) are shown following the subtraction of the systemic velocities from the three instruments (Table~\ref{RV-derived_param})} 
\end{figure}

\begin{table*}
\caption{RV-derived parameters of K2-19b and K2-19c from a two-planet model.}
\centering
\begin{tabular}{lc}
\hline
\hline
\noalign{\smallskip}
Parameter & Value \\
\hline
\noalign{\smallskip}
        & \textbf{K2-19b} \\
\noalign{\smallskip}
RV semi-amplitude variation K$_\mathrm{b}$ [\ms] ${^*}$ & $18.8\pm2.4$ \\
\noalign{\smallskip}
Eccentricity $e_\mathrm{b}$ ${^*}$ & $0.094\pm0.075$    \\
\noalign{\smallskip}
Argument of periapsis $\omega_\mathrm{*,b}$ [$\deg$] ${^*}$  & $100_{-70}^{+37}$\\
\noalign{\smallskip}
Epoch of periapsis T$_\mathrm{p,b}$ $^{**}$& $2\,456\,812.44\pm0.44$ \\
\noalign{\smallskip}
Planet mass M$_\mathrm{b}$ [M$_\oplus$] $^{**}$ & $ 54.8\pm 7.5$ \\
\noalign{\smallskip}
\noalign{\smallskip}
\noalign{\smallskip}
               &  \textbf{K2-19c} \\
\noalign{\smallskip}
RV semi-amplitude variation K$_\mathrm{c}$ [\ms] ${^*}$ &  $1.77_{-1.28}^{+2.26}$ \\     
\noalign{\smallskip}
Eccentricity $e_\mathrm{c}$ & 0 (fixed)    \\
\noalign{\smallskip}
Planet mass M$_\mathrm{c}$ [M$_\oplus$] $^{**}$ & $ 5.9_{-4.3}^{+7.6}$ \\
\noalign{\smallskip}
\noalign{\smallskip}
\noalign{\smallskip}
               & \textbf{Systemic RV} \\
\noalign{\smallskip}FIES systemic RV $\gamma_{\mathrm{FIES}}$ [\kms] ${^*}$ & $7.1951\pm0.0030$  \\
\noalign{\smallskip}
HARPS-N systemic RV $\gamma_{\mathrm{HARPS-N}}$ [\kms] ${^*}$& $7.3153\pm0.0019$  \\
\noalign{\smallskip}
HARPS-S systemic RV $\gamma_{\mathrm{HARPS}}$ [\kms] ${^*}$& $7.3272\pm0.0020$ \\
\noalign{\smallskip}
\hline
\label{RV-derived_param}
\end{tabular}

\tablefoot{

Orbital periods and epochs were input values taken from \citet{Armstrong2015}.\\ ~~\tablefoottext{$*$}{Direct output from RV fit.}\\ \tablefoottext{  $**$}{Parameter derived from fit-outputs.}
  }

\end{table*}

We fitted one-planet and two-planet Keplerian models to the FIES, HARPS-N, and HARPS RV data. In the first case, we assumed that the observed Doppler shift is caused \emph{entirely} by the largest transiting planet K2-19b; in the second case we assumed that both planets contribute to the observed RV variation. 

The RV analysis was done using \texttt{pyaneti}, a Python/Fortran software suite based on MCMC sampling (Barragan et al., in preparation). The code implements the ensemble sampler with affine invariance algorithm of \citet{Goodman2010}. It finds the best fitting parameters of the following equation RV\,$=\gamma_i + \sum_j^N K_j \left[ \cos(\theta_j + \omega_{\star,j}) + e_j \cos \omega_{\star,j} \right]$, where $\gamma_i$ are the systemic velocities as measured by the three instruments, $j$ refers to each planet, $N$ is the number of planets, $K_j$, $\theta_j$,  $e_j$ are the RV semi-amplitude variation, true anomaly, and orbit eccentricity of each planet $j$, respectively, and $\omega_{\star,j}$ the argument of periapsis of the star's orbit.

We fixed orbital periods and mid-times of first transit to the values given by \citet{Armstrong2015}, i.e., P$_\mathrm{orb,\,b}$=$7.919454_{-0.000078}^{+0.000081}$~days and T$_\mathrm{0,\,b}$=$2456813.38345_{-0.00039}^{+0.00036}$ (BJD$_\mathrm{TDB}$) for K2-19b, and P$_\mathrm{orb,\,c}$=$11.90701_{-0.00039}^{+0.00044}$~days and T$_\mathrm{0,\,c}$=2456817.2759$\pm$0.0012 (BJD$_\mathrm{TDB}$) for K2-19c. Using different ephemeris -- as those presented in Sect.~\ref{RV_TTV_Analysis} or those provided by \citet{Narita2015}, \citet{Barros2015}, \citet{Sinukoff2016} -- gives consistent results well within the error bars. 
 
For the eccentricity and the argument of periapsis we set uninformative uniform priors using the parametrization $ a = \sqrt{e} \sin \omega_{\star,\mathrm{b}}$ and $ b= \sqrt{e} \cos \omega_{\star,\mathrm{b}}$ with both a and b within the range $]-1,1[$, where the reversed brackets mean that the range endpoints are excluded. 

To ensure that e < 1, we also impose the condition $a^2 + b^2 < 1$, which was checked for all the iterations. 

For the systemic RVs, we set uniform priors in the  range $\gamma_i = [7.17,7.35]\, {\rm km\, s^{-1}}$, whereas the $K_\mathrm{b,c}$ were unconstrained, with initial values randomly set between $K_\mathrm{b,c} = [0.5,1000] \,{\rm m\,s^{-1}}$.
For the two-planet fit, we fixed $\sqrt{e_\mathrm{c}} \sin \omega_{\star,\mathrm{c}} = \sqrt{e_\mathrm{c}} \cos \omega_{\star,\mathrm{c}} = 0 $ and fit only for the RV semi-amplitude variation $K_\mathrm{c}$.
We evolved 1\,000 independent chains and ran 50\,000 additional iterations, with a thinning factor of 50, once convergence was reached. The final parameter estimates were obtained by combining the points from all the chains, leading to a total number of 10$^6$ points for each parameter. 

Assuming a stellar mass of M${_\star}$\,=\,0.918$\pm$0.064~\Msun\ (Sect.~\ref{StellarParameters}), modeling the RV data with one Keplerian orbit gives a mass of M$_\mathrm{b}$\,=\,58.6\,$\pm$\,4.6\,M${_\oplus}$ for K2-19b, with a chi-square value of 15.6. The two-planet modeling gives a similar value of M$_\mathrm{b}$\,=\,54.8\,$\pm$\,7.5\,M${_\oplus}$ for K2-19b, and a mass of M$_\mathrm{c}$\,=\,$5.9^{+7.6}_{-4.3}$\,M$_\oplus$ for K2-19c, with a chi-square value of 17.5. We conclude that the RV data do not allow us to significantly detect the Doppler reflex motion induced by K2-19c. Nevertheless, due to the knowledge about the presence of the two planets in this system, and the marginal $\sim$2-sigma RV detection of K2-19c, the two-planet fit is the preferred one. The parameter estimates -- defined as the median values of the posterior probability distributions -- are given in Table~\ref{RV-derived_param} along with the 68\,\% credible interval. Results will be further discussed in Sect.~\ref{RV_TTV_Analysis} and Sect.~\ref{Discussion}.
 
\section{Combined RV and transit timing analysis}
\label{RV_TTV_Analysis}

In a further analysis we derived masses and orbital parameters of K2-19b and K2-19c using the code \texttt{TRADES} \citep{Borsato2014} to simultaneously model RV measurements and TTV data. \texttt{TRADES} combines the Particle Swarm Optimization \citep[\texttt{PSO};][]{Tada2007} with the Levenberg-Marquardt algorithm \citep[\texttt{LM};][]{More1980}. We used our FIES, HARPS-N, and HARPS RVs (Sect.~\ref{RVs}) along with 20 transit mid-times (TTs) published by \citet{Narita2015}\footnote{We used only 20 TTs out of the 21 listed by \citet{Narita2015} because there are two TTs identified with epoch 34, observed with two different facilities, which we joined into a single data-point with a smaller uncertainty.}. The ground-based observations from \citet{Barros2015} were not used in our analysis since the authors do not list the transit mid-times, nor can the measurements be retrieved from their figures with sufficient precision. Considering the rather similar epochs and O$-$C times between the \citet{Barros2015} and \citet{Narita2015} follow-up transits, an inclusion of the \citet{Barros2015}'s transits is unlikely to cause significantly different results. Given the amplitude of the observed RV peak-to-peak variation ($\sim$40~\ms; Fig.~\ref{RV_Curve}), we set a very conservative range of 0$<$M$_\mathrm{p}$$<$100~M$_\oplus$ for the masses of the two planets. To account for the two degenerate solutions found by \citet{Narita2015}, we assumed a wide range for the orbital periods, i.e., P$_\mathrm{orb,\,b}$=7.8-8.0~days and P$_\mathrm{orb,\,c}$=11.5-12.5~days. We reduced the correlation between eccentricity $e$ and argument of periapsis $\omega_\mathrm{*}$ of the star\footnote{We note that TRADES uses internally the orbital elements of the planets; the discussion here and the values in Table 3 have been changed to the angle of periapsis of the central star, which is the habitually given value, with $\omega_\mathrm{*} = \omega_\mathrm{planet} \pm 180 \deg$ } by fitting instead for the combinations $e \cos \omega_\mathrm{*,i}$ and $e \sin \omega_\mathrm{*,i}$, where the index i refers to planet b and c. We limited the possible eccentricities to $e$$<$0.5, given the phased RV curve, and let the arguments of periapsis $\omega_\mathrm{*}$ and mean anomalies $\nu$ vary freely between $0$ and $360$ degrees. We used the orbital inclinations as given in \citet{Barros2015}. We fixed the longitudes of nodes of both planets to zero degrees. The stellar mass was left to float around the value with gaussian errors found in  Sect.~\ref{StellarParameters} (see Table 3).

\begin{figure*}
\centering
\includegraphics[width=\textwidth]{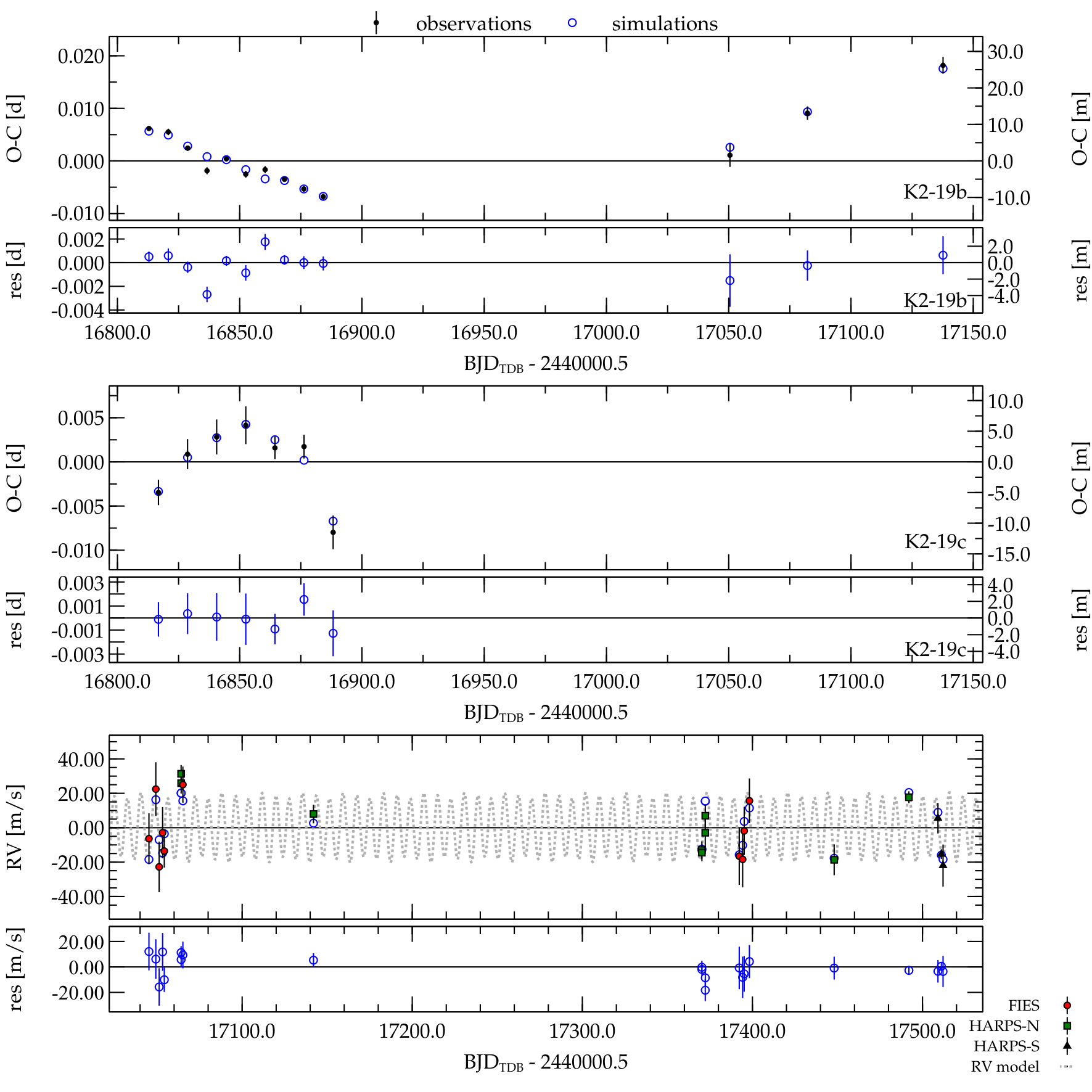}
\caption{\emph{Upper panels}: Observed-Calculated (O$-$C) times and residual plots for transit timings of K2-19b \emph{(top)} and K2-19c \emph{(middle panel)}.  O$-$C values have been computed by subtracting a linear ephemeris from each timing measurement (black dots), taken from K2 data and from ground-based follow-up by \citet{Narita2015}. Open blue circles indicate the best fitting model found by \texttt{TRADES}; the lower sub-panels indicate the residuals of the \texttt{TRADES}-model. \emph{Bottom panel}: Radial velocity measurements from FIES (red circles), HARPS-N (green squares) and HARPS-S (black triangles) as well as the best fitting model from TRADES (blue open circles). The gray dotted line shows TRADES' RV model of both planets across the observation time window. Residuals against the model are shown in the lower sub-panel.}
\label{fig:bestfit}
\end{figure*}

The PSO simulation evolved 250 initial orbital configurations for 15\,000 iterations. We used the best fitting solution as the initial guess for the \texttt{LM} algorithm. \texttt{TRADES} found a best-fit solution with a reduced chi-square $\chi^{2}_\textrm{red}$=1.57 (degrees of freedom $\mathrm{dof}$=32). The parameter estimates are listed in Table~\ref{ParameterTable} along with the confidence intervals at the $15.87^\mathrm{th}$ and $84.14^\mathrm{th}$ percentile of the residual distribution. The confidence intervals were computed with a bootstrap Monte-Carlo analysis running 2\,000 iterations and re-scaling the error bars by the quantity $\sqrt{\chi^{2}_\mathrm{r}}$ \citep{Bruntt2006, Southworth2007, Southworth2008}. By applying the frequency map analysis method \citep{Laskar1992}, we found that the derived orbital configuration is stable. Figure~\ref{fig:bestfit} shows the simulated data points from the best-fit solution overlaid on the observed data. The derived planet masses are M$_\mathrm{b}$=$ 54.4_{-9.5}^{+8.2}$~M${_\oplus}$ and M$_\mathrm{c}$=$7.5_{- 1.4}^{+3.0}$~M${_\oplus}$. We note that the chi-square of TRADES' RV-model (which is based on both RVs and TTVs) against the radial velocity data is 19.0, which is larger than the corresponding chi-square (17.5) from the RV-analysis presented in the previous section, and the planet masses from TRADES have also larger errors.

\section{Discussion}
\label{Discussion}

We present FIES, HARPS-N, and HARPS radial velocity follow-up observations of K2-19, a K0 dwarf star hosting two planets whose orbital periods are close to the 3:2 mean motion resonance. We aim at using high precision RV measurements to constrain the masses of K2-19b and K2-19c. From an analysis based only on our RV measurements and with the stellar parameters derived in Sect.~\ref{StellarParameters}, we estimate that K2-19b has a mass of M$_\mathrm{b}$\,=\,54.8\,$\pm$\,7.5~M${_\oplus}$ and K2-19c has a mass of M$_\mathrm{c}$\,=\,$5.9_{-4.3}^{+7.6}$~M${_\oplus}$. A combined analysis of RV and TTV measurements (Sect.~\ref{RV_TTV_Analysis}) resulted however in a slightly lower mass of M$_\mathrm{b}$=$54.4\pm8.9$~M$_\oplus$ for K2-19b and higher mass of M$_\mathrm{c}$=$7.5_{-1.4}^{+3.0}$~M$_\oplus$ for K2-19c. The two mass values of K2-19b are consistent with each other, deviating by less than one sigma. Regarding K2-19c, the analysis of the RVs on its own, without prior knowledge of the existence of planet \emph{c}, would not provide relevant evidence for its existence. Given that knowledge, its RV-signal was however marginally detected, implying a mass of  $5.9_{-4.3}^{+7.6}$~M$_\oplus$, whereas the combined RV and TTV analysis led to a similar but better constrained mass of $7.5_{-1.4}^{+3.0}$~M$_\oplus$. 

We note that the RV+TTV fits force radial velocity amplitudes that are smaller for planet~$b$ and larger for planet~$c$, in both cases by a similar amount of $\sim$1\,\ms\ relative to the RV-only fit. Considering the known difficulties to quantify the contribution from stellar activity to RV amplitudes on the \ms\ level, we expect that the RV amplitudes have larger uncertainties than those derived from the fits of Sect.~\ref{RV_Analysis} and~\ref{RV_TTV_Analysis}, which in both cases designate the RV signal to be entirely caused by the orbiting planets. Given this, the results from the two methods can be considered to be in agreement, implying however that no detection of planet~$c$ can be claimed from the radial velocities.

A possible concern is that the known period variation of K2-19b may affect the Keplerian fit to the RV data, which assumed constant periods. The maximum deviations from constant period during our three principal groups of RV observations, near BJDs ending in 7050, 7400 and 7500, are of 10, 70 and 90 minutes, respectively, based on \citet{Barros2015}'s TTV prediction for these epochs (their Fig.~5; with the last value for BJDs $\approx$7500 being an extrapolation). Such TTVs are not expected to significantly affect the Keplerian RV fit as they may cause only small shifts in K2-19b's phases of 0.0009, 0.006 and 0.008, respectively, implying RV variations of less than 1\,\ms. Therefore, the RV error bars of $\ga 6$~\ms\ for most of our measurements will dominate over RV deviations due to phase shifts as long as K2-19b's phases remain within 0.04 (or 8 hours) relative to phases derived from an ephemeris based on the mean period given in Table~\ref{ParameterTable}. A similar argument can be made for K2-19c, where maximum TTV's of 250 minutes can be predicted, corresponding to a maximum phase-shift of 0.015. Due to K2-19c's small RV amplitude, its RV values would be affected by such a phase shift only on the cm/s level. The RV fit should therefore not be affected by the known TTVs. 

Assuming a planet radius\footnote{Derived from the planet-to-star radius ratio of 0.0745\,$\pm$\,0.0010 for K2-19b and 0.0451\,$\pm$\,0.0007 for K2-19c from \citet{Barros2015} and our stellar radius estimate given in Table~\ref{ParameterTable}.} of 7.16\,$\pm$\,0.91~R${_\oplus}$ for K2-19b and of 4.34\,$\pm$\,0.55~R${_\oplus}$ for K2-19c, our estimate of the planets' masses from the combined analysis of RV and TTV measurements implies mean densities of $\rho_\mathrm{b}$=0.85\,$\pm$\,0.31~\gcm3  for planet $b$ and $\rho_\mathrm{c}$\,=\,$0.51_{-0.21}^{+0.27}$~\gcm3 for planet $c$. This density points to a likely gaseous planet with a dense core, similar to the conclusion by \citet{Barros2015}. For K2-19c, our derived radius and density would imply a planet somewhat larger than Neptune, but more lightweight,  without the silicate and nickel-iron core present in Neptune. 

Comparing our results to those of \citet{Barros2015}, the authors derived from their photo-dynamical analysis a mass of M$_\mathrm{b}$\,=\,$44\pm12$~M${_\oplus}$ for K2-19b and of M$_\mathrm{c}$\,=\,$15.9_{-2.8}^{+7.7}$~M${_\oplus}$ for K2-19c, with correspondingly lower respectively higher densities. In the case of K2-19b, our value estimated is larger than around 1-$\sigma$. The difference becomes even slightly larger if we adopt in our RV analysis the $\sim$3\,\% larger stellar mass of \citet{Barros2015}. For K2-19c, our mass determinations are significantly lower than the one from \citet{Barros2015}, with a difference of more than~2-$\sigma$.
 
Considering that the planet-masses of \citet{Barros2015} were essentially derived from TTV's, and that the masses from our combined RV+TTV analysis are between those from \citet{Barros2015} and our RV-only analysis, we suspect that the TTVs force the mass derivation of K2-19b towards lower values than given purely by RV data. This is also consistent with the fact that the errors of the planet masses (driven by the errors in the RV semi-amplitude $K$) in our RV-only solution are smaller than those from our combined analysis. In other words, the RV fit that is also driven by TTVs provides a chi-square (19.0) that is higher than the chi-square (17.5) of the RV fit obtained ignoring the TTVs. The sources for these divergences in RV and TTV results are difficult to establish. Uncorrected systematics in either the RV or the TTV data and their interpretation might account for the discrepancies between the two results. The relatively small difference among the chi-squared of the RV models versus the much larger difference in the mass-uncertainties of planets $b$ and $c$ indicates indeed an underestimation of the formal RV errors. We also note that our TTV analysis and the one by \citet{Barros2015} is not identical, with \citet{Barros2015} considering also the shapes of transits.

During the revision of this paper, a further study involving RV observations of K2-19 was published by \citet{Dai2016}. With an eccentric RV model, they obtained mass estimates that are inconsistent with our work, i.e., $31.8_{-7.0}^{+6.7}$~M$_\oplus$ for planet $b$ and $26.5_{-10.8}^{+9.8}$~M$_\oplus$ for planet $c$. A revision of the RV values in their Table 7 shows that the majority of their RV data were obtained with the Carnegie Planet Finder Spectrograph (PFS) on the 6.5-m Magellan/Clay Telescope. In most of their observing nights, they obtained 3 nightly RV points of K2-19. Most of these nightly groups show differences between individual data-points that are much larger than their quoted uncertainties of $\sim$5\,\ms, in many cases with nightly RV-variations exceeding 20\,\ms. These variations are too large to be ascribed to a physical origin in the K2-19 system and arise apparently from an unrecognized source of measurement errors. Given the strong nightly RV shifts in the PFS data, unrecognized error-sources that affect frequencies longer than single nights might be present as well, with potential effects onto the planets' RV amplitudes.

To date there are only few planets whose masses have been derived using both methods. As examples, we cite \citet{Nesvorny2013} and \citet{Barros2014} who derived the mass of the same planet, \object{Kepler-88c} (formerly known as \object{KOI-142c}) using TTVs and RV measurements, respectively. The first team measured a mass of 0.62\,$\pm$\,0.03 M$_\mathrm{Jup}$ from TTVs detected on the transiting planet \object{Kepler-88b}, from which they determined the mass of the non-transiting planet Kepler-88c. The presence of Kepler-88c was later confirmed by \citet{Barros2014} using RV measurements. They derived a mass of $0.76_{-0.16}^{+0.32}$ M$_\mathrm{Jup}$ for planet $c$, which agrees with TTV predictions of \citet{Nesvorny2013} and provides an independent validation of the TTV method.

In the Kepler-89, system, however \citet{Weiss2013} give RV derived masses for the planets Kepler-89$c$ and $d$, with a marginal detection of the more lightweight planet $c$. \citet{Masuda2013} and \citet{Hadden2016} present each a TTV analysis of the same system. They are able to determine the mass of planet $c$, found to be within the range of Weiss et al.'s RV measurement, whereas their mass-determinations of planet $d$ indicate a mass $\approx $3 times lower than that from RVs.

Another case is the reanalysis of RV and TTVs data of the \object{Kepler-9} planetary system by \citet{Borsato2014} using the same analysis tool (TRADES) as in this paper, and who estimated planet masses $\sim$55\,\% lower than reported by in the original discovery paper \citep{Holman2010}, which was based on the combination of RVs, transit times and durations. In this case, \citet{Borsato2014} ascribe the discrepancy to the longer Kepler light-curve analysed by them, as well as different approaches in the  interpretation of the TTVs.


Recently, \citet{Cubillos2017} study all Neptune-like planets for which both masses and radii are known, and also note that the planets measured by TTVs have typically lower densities. \citet{Weiss2014} already discuss possible causes of this difference, quoting systematic underestimations for masses from TTVs (e.g. from damping of TTVs by other planets) or selection effects that make lower-density planets more amenable to be detected by TTVs.
\citet{Cubillos2017} present also a hypothesis that the lower densities of TTV planets are possibly caused by high-altitude clouds or hazes that lead to inflated radii. This would however only apply if there are systematic differences in the radii of RV and TTV measured planets, caused by selection effects. In any case, this effect cannot account for any differences if masses from both RV and TTV measures are known.


From the results of this work, RV and TTV measurements complement each other, with slight tensions remaining. For planet {\it b}, which is the more significant detection in the K2-19 system, the addition of RV measurements raised the mass obtained previously from TTV's by Barros et al. (2015), but within error bars. Planet $c$ was barely detected in our RV data, whereas we know that it exists from the transits and it is detected with much higher significance from TTV data alone or from their combination with RV data. This difference in the detection quality of a low-mass planet arises most likely from the limited precision of the RV data. We also note that RV results from different teams may vary strongly, which may have its origin in unrecognized issues of their calibration. To resolve such tensions among different results, a better understanding about the causes that may generate systematics between RV and TTV methods, but also between results obtained by the same method are desirable. More mass-measures of planets with both RVs and TTV methods should also lead to a better understanding of the origins of such differences.

\onecolumn
\begin{table*}
\caption{K2-19 system parameters.}
\centering
\begin{tabular}{lr}
\hline
\hline
\noalign{\smallskip}
Parameter & Estimate \\ 
\noalign{\smallskip}
\hline
\noalign{\smallskip}
\multicolumn{2}{l}{\emph{Measured stellar parameters}} \\
\noalign{\smallskip}
Effective temperature \teff\ [K] & 5250$\pm$70 \\
\noalign{\smallskip}
Surface gravity $\log g_{*}$ [cgs] & 4.50$\pm$0.10  \\ 
\noalign{\smallskip}
Iron abundance [Fe/H] [dex] & 0.10$\pm$0.05  \\
\noalign{\smallskip}
Microturbulent velocity \vmic\ [\kms] & 0.8  \\
\noalign{\smallskip}
Macroturbulent velocity \vmac\ [\kms] & 2.5  \\
\noalign{\smallskip}
Projected rotational velocity $v$\ sin $i_*$\ [\kms] & 3.00$\pm$0.50  \\
\noalign{\smallskip}
Stellar rotation period $P_\mathrm{rot}$ [days] & 20.54$\pm$0.30  \\
\noalign{\smallskip}
\hline
\noalign{\smallskip}
\multicolumn{2}{l}{\emph{Derived stellar parameters}} \\
\noalign{\smallskip}
Star mass M${_*}$ [\Msun] & 0.918$\pm$0.064 \\
\noalign{\smallskip}
Star radius R${_*}$ [\Rsun] & 0.881$\pm$0.111 \\
\noalign{\smallskip}
\hline
\noalign{\smallskip}
\multicolumn{2}{l}{\emph{K2-19b}}\\
\noalign{\smallskip}
RV semi-amplitude variation K$_\mathrm{b}$ [\ms] ${^*}$ & $18.5\pm 3.0$ \\
\noalign{\smallskip}
Planet/Star mass ratio & $ 0.00018\pm0.00003$ \\
\noalign{\smallskip}
Planet mass M$_\mathrm{b}$ [M$_\oplus$]        & $ 54.4\pm8.9$ \\
\noalign{\smallskip}
Orbital period P$_\mathrm{orb,b}$ [days]       & $  7.91951_{- 0.00012}^{+0.00040}$ \\
\noalign{\smallskip}
$e_\mathrm{b}\cos \omega_\mathrm{*,b}$           & $ -0.0004_{- 0.0190}^{+0.0380}$ \\
\noalign{\smallskip}
$e_\mathrm{b}\sin \omega_\mathrm{*,b}$           & $  0.023_{- 0.23}^{+0.01}$ \\
\noalign{\smallskip}
Eccentricity $e_\textrm{b}$                    & $ 0.023_{-0.020}^{+0.240}$ \\
\noalign{\smallskip}
Argument of periapsis $\omega_\mathrm{*,b}$ [$\deg$] & $271\pm12$ \\
\noalign{\smallskip}
Epoch of periapsis T$_\mathrm{p,b}$& $2\,456\,809.5\pm0.1$ \\
\noalign{\smallskip}
\hline
\noalign{\smallskip}
\multicolumn{2}{l}{\emph{K2-19c}} \\
\noalign{\smallskip}
RV semi-amplitude variation K$_\mathrm{b}$ [\ms] ${^*}$ & $2.3_{-0.4}^{+0.9}$ \\
\noalign{\smallskip}
Planet/Star mass ratio & $ 0.000024_{-0.000005}^{+0.000010}$ \\
\noalign{\smallskip}
Planet mass M$_\mathrm{c}$ [M$_\oplus$]       & $  7.5_{-1.4}^{+3.0}$ \\
\noalign{\smallskip}   
Orbital period P$_\mathrm{orb,c}$ [days]      & $ 11.9066_{- 0.0014}^{+0.0021}$ \\
\noalign{\smallskip} 
$e_\mathrm{c}\cos \omega_\mathrm{*,c}$          & $  -0.0153_{- 0.0270}^{+0.0088}$    \\
\noalign{\smallskip}   
$e_\mathrm{c}\sin \omega_\mathrm{*,c}$          & $  0.1826_{- 0.2800}^{+0.0002}$    \\
\noalign{\smallskip}   
Eccentricity $e_\mathrm{c}$                   & $ 0.183_{-0.003}^{+0.283}$         \\
\noalign{\smallskip}   
Argument of periapsis $\omega_\mathrm{*,c}$       & $275\pm5$               \\
\noalign{\smallskip}  
Epoch of periapsis T$_\mathrm{p,c}$& $2\,456\,811.55\pm0.15$ \\
\noalign{\smallskip} 
\hline
\noalign{\smallskip}
\multicolumn{2}{l}{\emph{Systemic RV}} \\
\noalign{\smallskip}FIES systemic RV $\gamma_{\mathrm{FIES}}$ [\ms] & $7195.64\pm4.22$  \\
\noalign{\smallskip}
HARPS-N systemic RV $\gamma_{\mathrm{HARPS-N}}$ [\ms] & $7327.10\pm2.61$  \\
\noalign{\smallskip}
HARPS-S systemic RV $\gamma_{\mathrm{HARPS}}$ [\ms] & $7311.91\pm1.79$ \\
\noalign{\smallskip}
\hline
\label{ParameterTable}
\end{tabular}
\tablefoot{
The planet orbital parameter estimates refer to the reference time BJD$_\mathrm{TDB}$=2456813.0.
  }  
 
\end{table*}

\begin{acknowledgements}

We thank the referee, Kento Masuda for a careful revision and comments improving this article. This work is based on observations obtained with the Nordic Optical Telescope, operated jointly by Denmark, Finland, Iceland, Norway, and Sweden, and the HARPS-N spectrograph on the Italian Telescopio Nazionale Galileo (TNG), operated by the INAF - Fundacion Galileo Galilei. Both telescopes are on the island of La Palma in the Spanish Observatorio del Roque de Los Muchachos of the Instituto de Astrof\'\i sica de Canarias. Based also on observations made with the ESO-3.6-m telescope at La Silla Observatory under program ID 097.C-0948. The research leading to these results has received funding from the European Union Seventh Framework Programme (FP7/2013-2016) under grant agreement No. 312430 (OPTICON). DN acknowledges an FPI fellowship BES-2013-067287 and HJD acknowledges support by grant ESP2015-65712-C5-4-R, both from the Spanish Secretary of State for R\&D\&i (MINECO). 

\end{acknowledgements}

\end{document}